\documentclass{PoS}

\title{Observer Access to the Cherenkov Telescope Array}

\ShortTitle{Observer Access to CTA}

\author{\speaker{J. Kn\"odlseder}\\
        IRAP, Toulouse, France\\
        E-mail: \email{jknodlseder@irap.omp.eu}}
\author{V. Beckmann\\
        APC, Paris, France\\
        E-mail: \email{beckmann@apc.in2p3.fr}}
\author{C. Boisson\\
        LUTh, Paris, France\\
        E-mail: \email{catherine.boisson@obspm.fr}}
\author{S. Brau-Nogu\'e\\
        IRAP, Toulouse, France\\
        E-mail: \email{sylvie.brau-nogue@irap.omp.eu}}
\author{C. Deil\\
        MPIK, Heidelberg, Germany\\
        E-mail: \email{Christoph.Deil@mpi-hd.mpg.de}}
\author{B. Kh\'elifi\\
        APC, Paris, France\\
        E-mail: \email{khelifi@in2p3.fr}}
\author{M. Mayer\\
        Humboldt University, Berlin, Germany\\
        E-mail: \email{michael.mayer@physik.hu-berlin.de}}
\author{R. Walter\\
        ISDC, Versoix, Switzerland\\
        E-mail: \email{Roland.Walter@unige.ch}}

\abstract{
The Cherenkov Telescope Array (CTA), a ground-based facility for very-high-energy (VHE) gamma-ray
astronomy, will operate as an open observatory, serving a wide scientific community to explore and to
study the non-thermal universe.
Open community access is a novelty in this domain, putting a challenge on the implementation of services 
that make VHE gamma-ray astronomy as accessible as any other waveband.
We present here the design of the CTA Observer Access system that comprises support of scientific 
users, dissemination of data and software, tools for scientific analysis, and the system to submit 
observing proposals. 
We outline the scientific user workflows and provide the status of the current developments.
}

\FullConference{The 34th International Cosmic Ray Conference,\\
		30 July- 6 August, 2015\\
		The Hague, The Netherlands}

\begin{document}

\section{Introduction}

The Cherenkov Telescope Array (CTA) will be the first very-high-energy (VHE) gamma-ray astronomy
facility that will be operated as a proposal-driven open observatory.
Access to observing time will be granted on the basis of peer-reviewed observation proposals that
respond to regular Announcements of Opportunity.
Allocation of observation time will be driven by the scientific excellence of the proposals.
The execution of observations will take place in service mode, in which observations fully pre-defined
by the users will be executed under suitable external conditions; the user can request specific 
telescope configurations, and specify a minimal sky quality for observations to be carried out.
The observatory will calibrate the acquired data using standard pipelines, and reconstruct photon 
directions and energies, allowing for several levels of quality selection.
Event lists, relevant instrument response functions (IRFs), and auxiliary data will be delivered
to the user in FITS format, together with a dedicated software package for data analysis.
After a proprietary period of one year, during which users granted observation time will have exclusive
access to data, the data will be made publicly and openly available in the CTA data archive.
High-level data, such as sky maps, spectra, light curves, and source catalogues, will be distributed
using Virtual Observatory (VO) protocols.
The distribution of event lists through VO protocols is under investigation.

Observer Access is one of the five work packages of the CTA Data Management work package
and provides all services and tools that are needed by Guest Observers
and CTA Archive Users to perform a successful scientific analysis of CTA data.
This paper provides a description of the Observer Access work package and its deliverables, 
which are mainly software and documents. 
Section \ref{sec:workflow} introduces the overall work flow for CTA science users,
section \ref{sec:products} describes the deliverables of the Observer Access work package, and
section \ref{sec:software} presents their design and status.
We conclude in section \ref{sec:conclusions}.

\section{Work Flow}
\label{sec:workflow}

Guest Observers will start with writing and submitting Guest Observer proposals to CTA in response
to an Announcement of Opportunity.
In preparing their proposals, Guest Observers will make use of calculators that provide information
about array sensitivity, target visibility, or required exposure time.
Once submitted, Guest Observer proposals will be evaluated by a Time Allocation Committee (TAC) 
that has been appointed by the CTA Observatory.
The Guest Observers will be informed about the outcome of this process, and successful proposals
will be scheduled for observation.
Guest Observers will be able to follow the status of the observations online, including the
scheduling, the data acquisition, the data processing, the data distribution and the ingestion of
the data in the public archive after the end of the proprietary period.
Guest Observers will download the processed data as well as the software tools that are necessary for
the scientific analysis from the web.
Science analysis will be performed on the Guest Observers' own computing infrastructure.
Web-based information about the data and the analysis software, comprising user manuals, cook 
books, etc. will also be available.
Eventually, Guest Observers will have attended dedicated CTA analysis schools or hands-on
workshops to familiarise themselves with the CTA Observatory, the CTA data, and the CTA analysis 
software.
If further assistance is needed, Guest Observers can contact a help desk (by web or by e-mail).

\begin{figure*}[!t]
 \center
 \includegraphics[width=15.2cm]{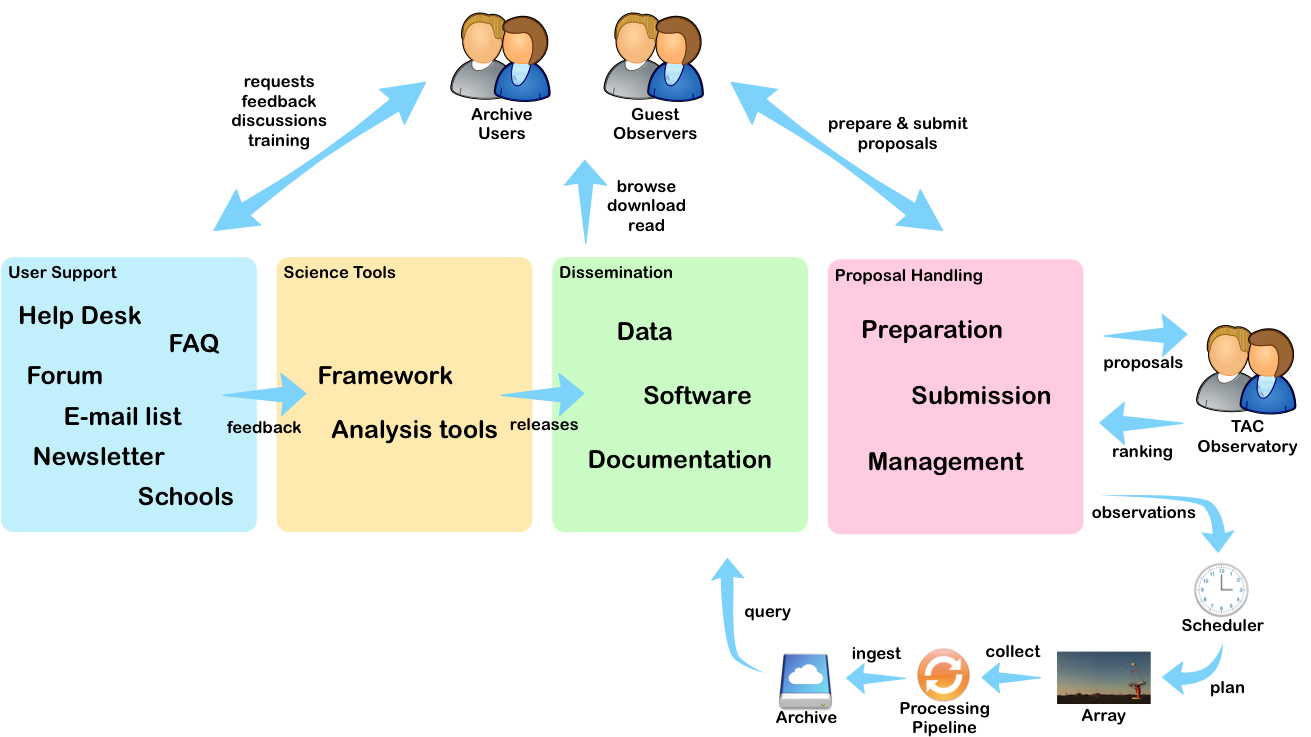}
 \caption{Overview over the Observer Access work flow and main components.}
 \label{fig:dataflow}
\end{figure*}

Archive Users will browse the public archive to access CTA data.
They either may download existing event data based on specific selection criteria (source location,
observation time interval, observation condition criteria, etc.), or they may ``prune'' the public
CTA event database to extract a specific dataset that meets the needs (e.g., specific event class
selection, energy range, etc.).
CTA images, spectra, light curves and event lists can also be accessed through Virtual Observatory 
(VO) protocols.
Archive Users will also download the analysis software, access any supporting material
as needed, and analyse the data on their own computing infrastructure.
Archive Users may also have attended dedicated CTA analysis schools or hands-on workshops
to familiarise themselves with CTA, the data and the software.
Archive Users may of course also contact the help desk to ask for assistance.

Figure \ref{fig:dataflow} illustrates the overall work flow for Guest Observers and Archive Users.
The figure distinguishes the four components that are needed to implement the services
and tools in this work flow: User Support, Science Tools, Dissemination and Proposal Handling.
Also shown are elements that lie outside the Observer Access work package, and that comprise
proposal evaluation by the TAC, scheduling, collecting, processing and archiving of the data.

\section{Observer Access Products}
\label{sec:products}

\begin{figure*}[!t]
 \center
 \includegraphics[width=15.2cm]{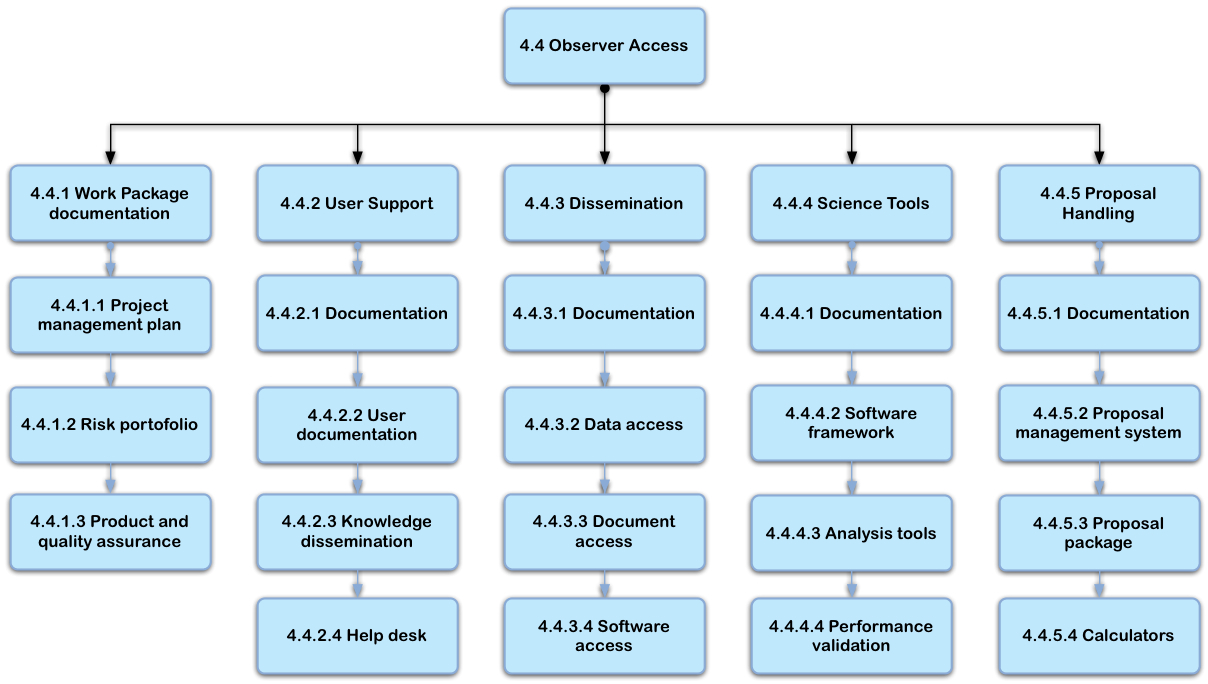}
 \caption{Observer Access Product Breakdown Structure (PBS).}
 \label{fig:oapbs}
\end{figure*}

The first two levels of the Observer Access work package Product Breakdown Structure (PBS) is 
represented in Fig.~\ref{fig:oapbs}.
Besides work package documentation, the Observer Access work package comprises four
products.
The User Support product provides the elements that are needed to support the scientific
users of CTA.
This comprises user documentation (user manuals, cookbooks, FAQ), knowledge dissemination 
(forum and e-mail list, newsletters, organisation of analysis schools), as well as a human-staffed 
help desk.
The Dissemination product provides all software that is needed to 
access data (data access),
documentation (document access), and
software (software access).
The Science Tools product provides the software that enables the user to derive images, spectra 
and light curves from calibrated, reconstructed and background reduced CTA event data.
The software is composed of a software framework and a set of analysis tools.
Extensive performance validation in the form of Science Challenges will be organised within the
CTA Consortium to assure the scientific quality of the software.
The Proposal Handling product comprises services and tools that are used by Guest
Observers to prepare (calculators) and to submit (proposal package) observation proposals.
In addition, services and tools are provided to the CTA Observatory for proposal management 
(proposal management system).
The proposal management system will also allow to follow the status of proposal acceptance, 
the observation execution progress, and the availability of the observation data.

\section{Design and Status}
\label{sec:software}

\subsection{Proposal Handling}

The Proposal Handling component is comprised of calculators, a proposal package and the
proposal management system.
The calculators include
a target visibility estimator (allowing to assess the visibility -- in terms of epoch in the year, 
duration, zenith angle -- of a potential CTA target from a given observatory site),
a sensitivity estimator (allowing to determine the sensitivity -- in terms of detectable flux --
for a given array configuration as function of observing conditions), and
an exposure time calculator (allowing to compute the needed exposure time for a given
target with specified flux as function of observing conditions).
The calculators will make use of parametrised or tabulated functions to provide a fast but reasonably
precise response to the Guest Observer, and will also serve as a reference for proposal evaluation
to the TAC.

The proposal package will comprise a web interface with web forms to provide proposal
information (e.g., name and affiliation of Principal Investigator, target name and coordinates,
requested array and configuration, executive summary, etc.).
It also includes a template for the scientific justification that will accompany the proposal.

All proposals will be handled by a dedicated Proposal Management System.
The system will comprise a proposal database, a proposal database management
system, and a system allowing Guest Observers to retrieve the status of their proposals,
including information about proposal evaluation, scheduling, data availability and 
release of the data into the public archive after the end of the proprietary period.
Proposals submitted through the web interface will be ingested by the Proposal
Management System.
The CTA Observatory, through its TAC, will use the system to access and rank the proposals.
Approved proposals will be fed by the Proposal Management System to the long-term observation
scheduler.
The Proposal Management System will then gather information about executed observations
from the scheduler, data processing information from the analysis pipeline, and data availability
and public release information from the data archive.
The Guest Observer will be able to access this information through a web interface.

\begin{figure*}[!t]
 \center
 \includegraphics[width=15.2cm]{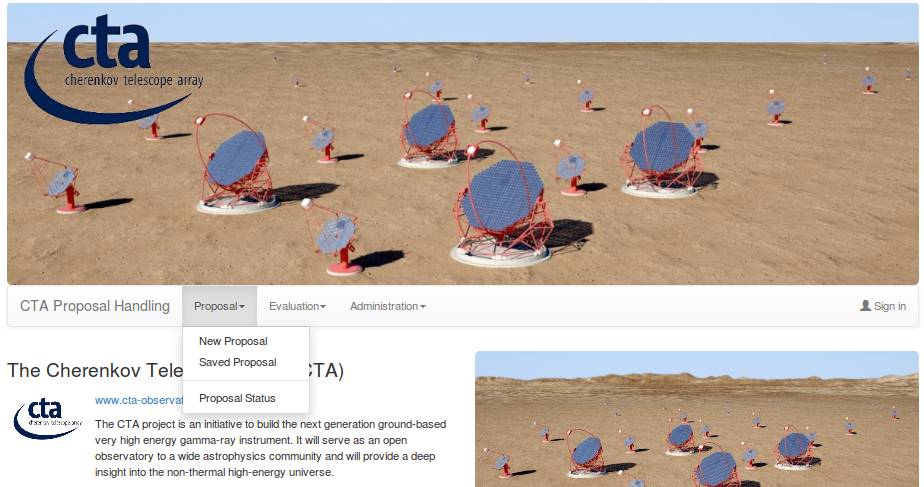}
 \caption{Prototype of the proposal management system.}
 \label{fig:proposal}
\end{figure*}

A screenshot for a prototype of the Proposal Management System is shown in Fig.~\ref{fig:proposal}.
This prototype is a custom development for CTA and is based on the Django framework, making use
of the jQuery and Bootstrap packages.
This technology is used also by the dissemination prototype (see next section), allowing an efficient
development path both in term of knowledge and manpower needs.
The Proposal Management System will be embedded in the CTA science gateway that serves as a
unique access point to CTA related information.

\subsection{Dissemination}

\begin{figure*}[!t]
 \center
 \includegraphics[width=15.2cm]{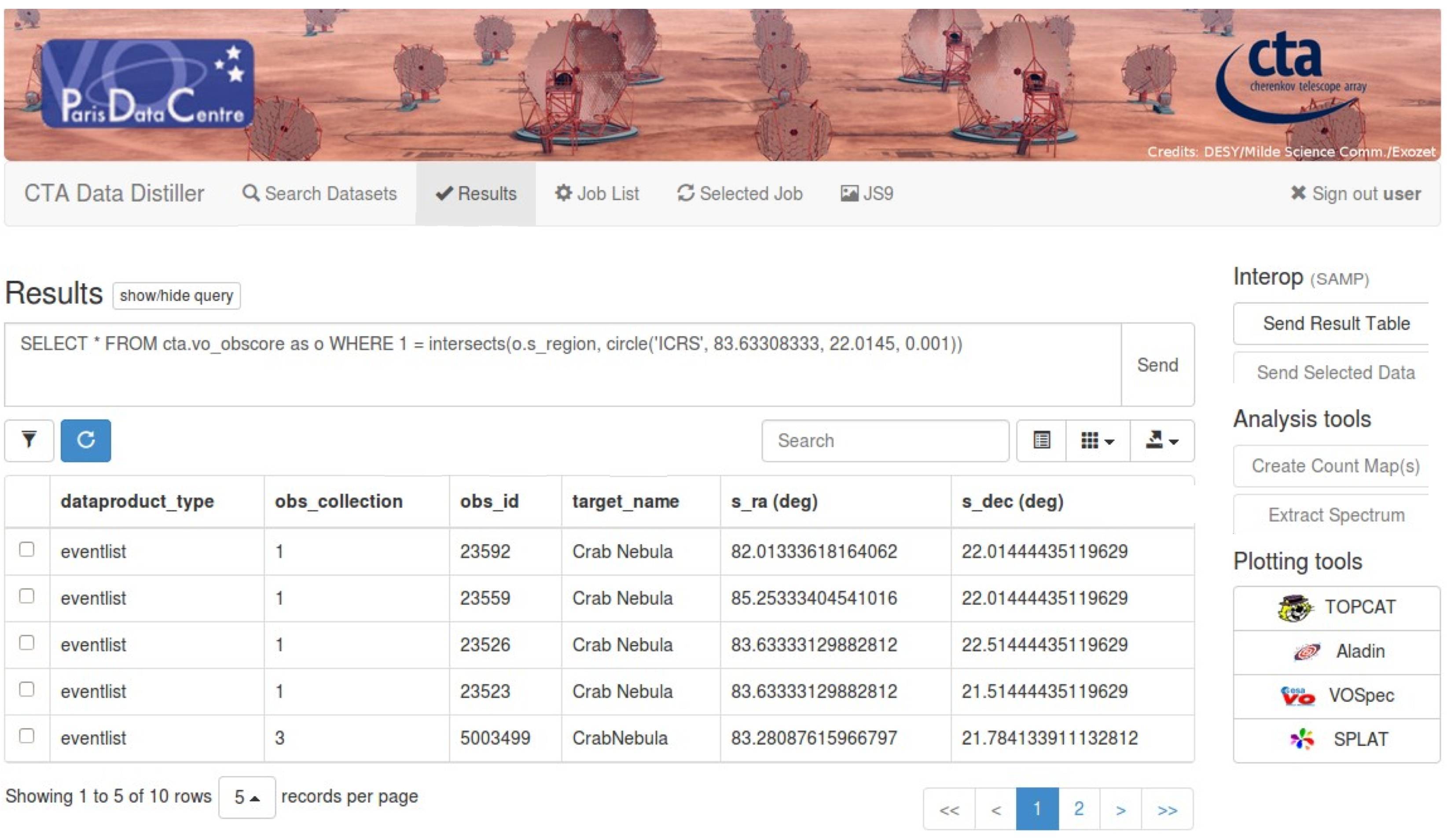}
 \caption{Prototype of the data access system using Virtual Observatory solutions.}
 \label{fig:access}
\end{figure*}

The Dissemination component provides all software that is needed to access data, software and
documentation.
The data access system enables the user to browse the CTA archive and to retrieve some data.
Selection of data sets of different categories will be allowed with various criteria (e.g., pointing
position, observation epoch, data quality, etc.).
The possibility to adapt an existing Browse System (such as HEASARC's Xamin) is under study,
but also a custom development that could be embedded in a VO environment is under
investigation.
Figure \ref{fig:access} shows a screenshot for a prototype of such a Browse System that is
developed using the Django, jQuery and Bootstrap frameworks.
Such a system, embedded in the CTA Science Gateway, would provide access to predefined 
CTA event data, but would also allow to process these data with some analysis tools and 
visualise the data using VO plotting tools.
In addition to the Browse System, the development of an Event Pruner is being considered.
This Event Pruner would give the user the possibility to retrieve customised event lists built
according to user defined criteria (sky position, time range, energy range, cuts, 
hadron/gamma/electron-ess).

In addition to event lists, high-level products such as images, light curves, spectra, and catalogues
will be distributed through a dedicated VO Data Server.
This will provide a standard access to the CTA archive and provide an easy mechanism for discovering 
data, data mining and cross correlation of multi spectral informations. 
It is also envisaged to generate high-level VO compliant data products on the fly according to
user specified wishes and source models.

\subsection{Science Tools}

\begin{figure*}[!t]
 \center
 \includegraphics[width=15.2cm]{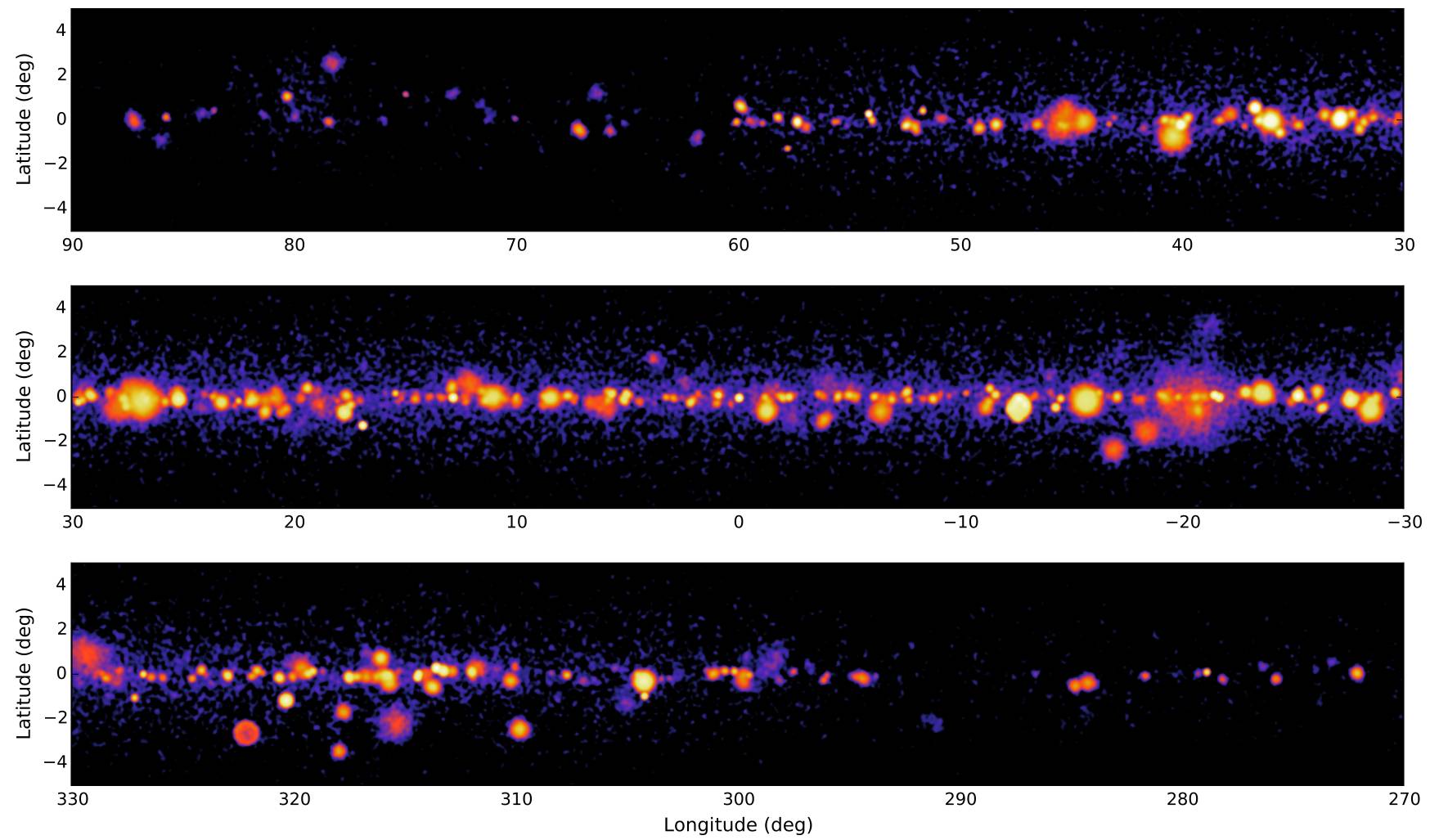}
 \caption{{\em ctools} simulation of the CTA Galactic plane survey.}
 \label{fig:simulations}
\end{figure*}

The Science Tools are the software that will be used by Guest Observers or Archive Users for
the science analysis of CTA data.
The software package will include tools to derive images, spectra and light curves from processed 
CTA high-level event data.
The Science Tools are primarily intended to be used by the CTA user community in their proper
computing environments, although they will also be embedded in the on-site and off-site
CTA analysis pipelines for production of standard high-level products.
The software will be distributed as binary packages for Linux and Mac OS X platforms,
and as source code for customised installations.

The Science Tools will be implemented as a custom software product for CTA, written in
C++ and interfaced to Python to support building of analysis scripts.
The software will consist of a software framework that provides the building blocks for gamma-ray 
data analysis, and of analysis tools that will implement the functional blocks that are required to 
analyse CTA data.
To guarantee an easy access to CTA data analysis for non-expert users, the Science Tools
will implement common astronomical standards and will follow the philosophy of tools developed
for other astronomical observatories (Fermi/LAT, INTEGRAL, XMM, Chandra, etc.).
In addition, the Science Tools will be compliant with standards defined under the International
Virtual Observatory Alliance (IVOA), enabling their interoperability with other IVOA tools 
(e.g.~Aladin, TOPCAT, CASSIS, VOSpec).

We have started the development of the {\em ctools} software package that is based on the
{\em\nobreak GammaLib} library as software framework and that we propose as the Science Tools for
CTA (see\break http://cta.irap.omp.eu/ctools/).
{\em ctools} and {\em GammaLib} are open source codes that are developed by a growing
community of developers.
The software is already widely used within the CTA Consortium for science simulations,
as illustrated in Fig.~\ref{fig:simulations}.
The software is also currently tested using data from the existing Cherenkov telescopes
H.E.S.S. and VERITAS.

The performance of the Science Tools will be validated using Science Challenges.
In each Science Challenge, simulated high-level science data will be made available
covering all major science topics and possible observations of CTA.
The CTA Consortium (and eventually even a larger user community) will then use the current 
version of the Science Tools to assess the performance of the system with respect to the 
expected science goals.
This includes aspects such as the easiness of performing science analysis, the flexibility of the
tools, the performance of the tools concerning accuracy of science results, achieved sensitivity,
and computing time.

\subsection{User Support}

The User Support component comprises the writing of all user documents that will support
the scientific CTA user in preparing and submitting proposals, getting data and software, and
performing the scientific analysis of the data.
In addition, user training in the form of hands-on schools and courses will be organised, and a 
users forum, a users e-mail list, and a CTA newsletter will assure the communication between the 
CTA Observatory and its users.
The CTA Observatory will also feature a help desk that will handle individual questions from
users and supports them in preparing and submitting observing proposals, retrieving CTA data,
and performing the science analysis.
Requests should be handled within a maximum delay of 1-2 working days.
Based on the requests received, the user documentation and tutorials will be improved, and
a list of Frequently Asked Questions (FAQs) will be compiled.

\section{Conclusions}
\label{sec:conclusions}

The Cherenkov Telescope Array is a unique observatory for very high-energy gamma-ray astronomy
that will benefit science in the fields of astronomy and astrophysics, particle physics and astroparticle
physics, cosmology, plasma physics and fundamental physics, by providing observers worldwide with
unprecedented data on astrophysical objects over an extensive range of gamma-ray energies.
Community access to CTA data is a backbone of the observatory, and the Observer Access work 
package will provide the services and tools that are necessary to grant a seamless and easy
access to the facility and its data.
Following the successful model of other observatory-type infrastructures, the Observer Access
work package will implement web-based proposal handling and data dissemination systems, 
provide easy-to-use open source tools for data analysis, and support users throughout the full
cycle from applying for observing time to analysing the data.
Observer Access is vital to the success of the CTA Observatory, and it will enable the VHE domain
to become an integral part of modern astronomy.

%

\end{document}